\documentclass[12pt,preprint]{aastex}

\newcommand{\myemail}{pollack@ucolick.org}
\newcommand{\etal}{\rm et~al.\ }

\slugcomment{{\sc Accepted to ApJ:} January 15, 2007}

\shorttitle{Circumnuclear Star Clusters in NGC 6240}
\shortauthors{Pollack, L. K., Max, C. E., \& Schneider, G.}

\begin{document}


\title{Circumnuclear Star Clusters in the Galaxy Merger NGC 6240, \\ Observed with Keck Adaptive Optics and HST}


\author{L. K. Pollack\altaffilmark{1}, and C. E. Max\altaffilmark{2}}
\affil{Department of Astronomy \& Astrophysics, University of California, Santa Cruz 95064}

\and

\author{G. Schneider\altaffilmark{3}}
\affil{Steward Observatory, The University of Arizona}

\altaffiltext{1}{email: \myemail}
\altaffiltext{2}{Center for Adaptive Optics, University of California Observatories/Lick Observatories, email: max@ucolick.org}
\altaffiltext{3}{email: gschneider@as.arizona.edu}


\begin{abstract}
We discuss images of the central $\sim$10 kpc (in projection) of the galaxy merger NGC 6240 at H and K' bands, taken with the NIRC2 narrow camera on Keck II using natural guide star adaptive optics.  We detect 28 star clusters in the NIRC2 images, of which only 7 can be seen in the similar-spatial-resolution, archival WFPC2 Planetary Camera data at either B or I bands.  Combining the NIRC2 narrow camera pointings with wider NICMOS NIC2 images taken with the F110W, F160W, and F222M filters, we identify a total of 32 clusters that are detected in at least one of these 5 infrared ($\lambda_{\rm c} > 1 \ \mu \rm m$) bandpasses.  By comparing to instantaneous burst, stellar population synthesis models \citep{bru03}, we estimate that most of the clusters are consistent with being $\sim$15 Myr old and have photometric masses ranging from $7 \times 10^{5} \rm M_{\odot}$ to $4 \times 10^{7} \rm M_{\odot}$.  The total contribution to the star formation rate (SFR) from these clusters is approximately $10 \rm M_{\odot} \ yr^{-1}$, or $\sim$10\% of the total SFR in the nuclear region.  We use these newly discovered clusters to estimate the extinction toward NGC 6240's double nuclei, and find values of $\rm A_{V}$ as high as 14 magnitudes along some sightlines, with an average extinction of $\rm A_{V} \sim 7$ mag toward sightlines within $\sim3$\arcsec  of the double nuclei.
\end{abstract}





\section{INTRODUCTION}\label{Introduction}
Ultraluminous Infrared Galaxies (ULIRGs) output more energy in the infrared than at all other wavelengths combined, and have $\rm L_{IR}>10^{12} L_{\odot}$. In the local universe, these relatively rare objects are frequently associated with peculiar galaxies, thought to be merging systems \citep{san88}.  Their high infrared emission comes from warm dust, heated by a combination of starbursts and active galactic nuclei (AGN) which are thought to be fueled by huge quantities of molecular gas that fall inward toward the nucleus during a major merger of two gas-rich spiral galaxies \citep{san96}.  While at the present epoch ULIRGs are rare in number, numerical simulations by \cite{mih04} suggest that they are not merely anomalies resulting from ``special" merger conditions, but rather their rarity is due to the fact that they represent a short-lived yet common phase of galactic evolution, namely when spiral galaxies collide to form massive elliptical galaxies.  Detections by the SCUBA telescope of hundreds of high-redshift sub-mm galaxies (SMGs) lend further support to this theory \citep[e.g.][]{bor03}.  The properties of SMGs are similar to those of local ULIRGs \citep{con03}, and the number density of SMGs is similar to the number density of massive elliptical galaxies at the present epoch.  Thus, assuming that structure in our Universe forms by hierarchical assembly, ULIRGs offer us the unique opportunity to glimpse a brief but important stage in a galaxy's life, when massive spirals transform into ellipticals.

Though NGC 6240 ($z=0.024$, $d=98$ Mpc, $1\arcsec=477$ pc; $H_{\circ}=71$ km $s^{-1}$ Mpc$^{-1}$, $\Omega_{\rm M}=0.27$, $\Omega_{\Lambda}=0.73$) technically lies at the high-luminosity edge of the LIRG class with $\rm L_{IR} \sim 6 \times 10^{11} \ L_{\odot}$, in many ways it is a prototypical ULIRG, and is commonly referred to as such.  NGC 6240 is a late-stage merger of two massive disk galaxies \citep{tacc99}; it has extended tidal tails and a close nuclear pair separated by  $\sim$1.5\arcsec, or a projected distance of 715 pc \citep[e.g.][]{max05, max06, bes01}.  Its K' band light is dominated by red supergiants in $\sim$20 Myr-old starbursts \citep{tec00}, and at the double nucleus two AGN have been discovered in hard X-rays \citep{kom03} as well as at 5 GHz \citep{gal04}.  As in other merging systems, most notably NGC 4038/9 (the Antennae galaxy) \citep{whi95}, star clusters have previously been observed in NGC 6240's main body and tidal tails with the WFPC2 F450W, F547W, and F814W filters \citep{pas03}.  But recently \citet{max05} detected previously unseen circumnuclear star clusters at J, H and K' bands using natural guide star (NGS) adaptive optics (AO) with the NIRC2 wide-field camera on Keck II.  These clusters, and other similar clusters in ULIRGs and galaxy mergers, may hold clues regarding a variety of open questions, from the distribution of dust and star formation in ULIRGs, to the formation and evolution of globular clusters (GCs).

Historically, measuring star formation rates (SFRs) in galaxy mergers has been difficult, and until recently the contribution of SMGs to the total luminous energy density in the high-redshift universe has been significantly underestimated.  A variety of heterogeneous techniques have been used to measure SFRs \citep{ken98}, but in starburst galaxies these are complicated by the presence of dust \citep{cal01}, including the usual methods of measuring SFRs through H$\alpha$ line fluxes or UV continuum luminosities.  In these cases astronomers have turned to far-infrared (FIR) wavelengths to measure the dust-reprocessed light.  But a well-known problem arises when using FIR luminosities to estimate SFRs in ULIRGs: in addition to young OB stars, AGN also help heat the dust.  While studies have shown that the luminosities of most ULIRGs are dominated by star formation even when AGN are present \citep{gen98, kew01}, accurately measuring the fractional AGN contribution to the bolometric luminosity ($L_{\rm bol}$) has proved difficult.  For example, \citet{arm06} reported on their detection of the [Ne {\sc v}] 14.3 $\mu$m emission line feature, with a flux of $5 \times 10^{-14}$ erg cm$^{-2}$ s$^{-1}$, which provides direct evidence of the buried AGN in NGC 6240.  They note that the [Ne {\sc v}]/[Ne {\sc ii}] ratio is often incorrectly treated as a reddening independent quantity, even though [Ne {\sc v}] comes directly from gas heated by the AGN, while [Ne {\sc ii}] can be dominated by a starburst.  Without correcting for dust, \citet{arm06} state that the large equivalent width of the $6.2\ \mu \rm m$ polycyclic aromatic hydrocarbon (PAH) emission feature, and the small [Ne {\sc v}]/[Ne {\sc ii}] and [Ne {\sc v}]/IR flux ratios, are consistent with an apparent AGN contribution of only $3-5$\% of $L_{\rm bol}$.  But after correcting the measured [Ne {\sc v}] flux for extinction using values implied by their SED fitting, \citet{arm06} estimate that the intrinsic fractional AGN contribution to the bolometric luminosity of NGC 6240 is much higher, between 20 and 24\%.

By detecting individual, massive, star clusters in merging galaxies we can trace the most violent star formation activity, while circumventing some of these hurdles which make accurately measuring SFRs a difficult task.  For example, the circumnuclear locations of clusters makes them easily distinguishable from AGN.  In addition, while the exact size distribution of dust grains in both star clusters and diffuse stellar populations is impossible to know, the complicated dust mixing models required for analyzing diffuse stellar fields reduce to simple dust screens toward clusters, if we assume that all intracluster dust has been evacuated.  Even when some intracluster dust is present, the dust screen approximation is still reasonable assuming that the amount of dust internal to the cluster is far less than the intervening dust along the line of sight through the entire merging galaxy.  Furthermore, individual clusters are well approximated by single stellar populations, making stellar population analysis of clusters more straightforward than for diffuse stellar fields.  Finally, near-infrared (NIR) cluster observations provide a more localized measure of the SFR compared to estimates based on longer wavelength observations.  But unfortunately, while star clusters seem to represent a common mode of star formation in spirals and LIRGs alike \citep[e.g.][]{lar02, sco00, fal05}, detecting dust-reddened clusters is challenging because longer wavelength observations are innately of lower spatial resolution.  However, with the improvement of AO systems on large, ground based telescopes, the possibility of detecting individual infrared star clusters, even at $z\sim.02$, is now a reality.

In order to measure \emph{localized} SFRs\footnote{We note that this method of using star clusters to measure localized SFRs, while beneficial in that it sidesteps complications due to AGN and dust, by no means traces the total SFR.  We preferentially trace star formation in the most massive, least dusty environments.  Less massive or more obscured clusters likely remain undetected in this work.} with fewer complications due to AGN and dust, we have obtained $\sim .04$\arcsec \ resolution images of roughly 10 kpc surrounding the merging galaxies' nuclei using the NIRC2 narrow-field camera with NGS AO at Keck.  We combine our data with HST images taken with the NICMOS NIC2 camera (using the F110W, F160W, and F222M filters) and the WFPC2 Planetary Camera (using the F450W and F814W filters).  Using the combined data set we study a total of 32 clusters with inhomogeneous wavelength coverage, partially due to variations in the fields of view, and partially due to non-detections in some wavebands.  In order to estimate the SFR associated with these clusters 
we estimate masses, ages, and extinctions toward these 32 clusters by comparing the clusters' spectral energy distributions (SEDs) to model SEDs from stellar population synthesis codes \citep{bru03}, employing simple, screen extinction models by \citet{dra03}.  Given the data currently available, we do not attempt to quantify the cluster's properties with the precision of previous studies of nearby mergers.  This work is meant to be a preliminary analysis of NGC 6240's recently discovered cluster population -- one of the most distant such populations ever detected; future high-resolution spectral observations will provide a much clearer understanding of the cluster's properties.  In \S\ref{Observations} we describe our observations, in \S\ref{Data Processing} we describe our data reduction, including the techniques we used to identify clusters and perform photometric calibrations, and in \S\ref{Results} and \ref{Discussion} we present and discuss our results.

\section{OBSERVATIONS}\label{Observations}

\subsection{Adaptive Optics at Keck}
We observed NGC 6240 with the NGS AO system on the Keck II telescope using the narrow-field camera (0.009942\arcsec \ pixel$^{-1}$; 4.74 pc pixel$^{-1}$) of the NIRC2 infrared instrument at H ($\lambda_{\rm c}=1.633 \mu$m) and K' ($\lambda_{\rm c}=2.124 \mu$m) bands.  The H and K' observations were obtained on 24 April 2005 and 17 August 2003, respectively.  The natural guide star (B = 13.5) is USNO-B 1.0 0924-0386013, located 35.8\arcsec \ east-northeast of NGC 6240's dual nuclei.

Due to the faint nature of the star clusters superposed on the diffuse background light from the galaxy, it is problematic to measure the resolution and Strehl ratio (S) at the location of the clusters, i.e. off-axis from the guide star.  We did not observe PSF star pairs and therefore cannot reasonably guess what the Strehl may be at the clusters' locations.  If there were no tip-tilt errors then the core of the off-axis point spread function (PSF) would be diffraction-limited, and at 2.1$\mu$m the full-width-half-maximum (FWHM) of this diffraction-limited, off-axis PSF core is $\sim 0.04\arcsec$ ($\sim 19$ pc at $z=.024$), making our observations slightly oversampled.  Throughout this paper we assume that the off-axis PSF and Strehl -- whatever they may be -- are constant across the off-axis FOV.  This assumption is well justified when a small FOV ($\sim10\arcsec$) is far off-axis ($\geq$30\arcsec), and when the areas of interest in the FOV are oriented perpendicular to the direction of the guide star.  Most of the clusters in our images are located along the north-south orientated double nuclei, and therefore are oriented approximately perpendicular to the direction of the guide star.  The assumed constancy of the resolution and Strehl ratio across the FOV implies that our detection limits and photometric measurements should be accurate in the relative sense; in \S\ref{caseII} (case II) we describe the steps we have taken to ensure that our photometric measurements are accurate in the absolute sense as well, by sidestepping the need to know the absolute Strehl ratio to perform photometry.

In Table~\ref{tableObsKeck} we outline the details of these Keck observations, including the on-axis Strehls achieved for the guide star on each night.  While the off-axis Strehl is unknown, the Strehl measured on the guide star is an indicator of the overall performance of the AO system, including the effects of wind turbulence, sampling rate, and weather.

\subsection{Archival HST Observations}\label{HSTObs}
We have analyzed archival WFPC2 images taken as part of the GO proposal 6430 (PI: R. van der Marel).  The observations were taken with the Planetary Camera (.0455\arcsec \ pixel$^{-1}$; 21.7 pc pixel$^{-1}$) using the F450W and F814W filters on 28 March 1999.  The theoretical FWHM of the diffraction-limited PSFs for these two WFPC2 images are $\sim.038\arcsec$ and $\sim.069\arcsec$, respectively.  At NGC 6240's distance, these angular resolutions correspond to 18.1 and 32.7 pc, respectively. The F450W and F814W filters are roughly equivalent to the Johnson-Cousins B and I filters, therefore throughout this paper we use the shorter Johnson-Cousins \emph{notation} interchangeably with the true WFPC2 filter names.  However, when performing all calculations and measurements we use the exact filter transmission functions for the WFPC2 filters.

\begin{figure}[! ht]
\epsscale{0.8}
\plotone{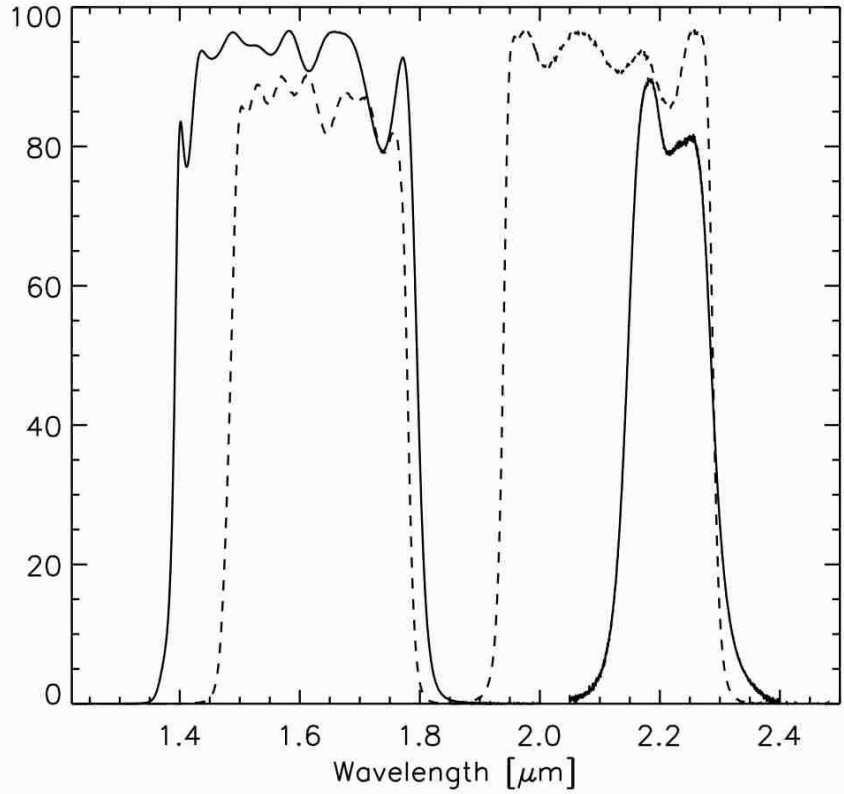}
\caption{Filter transmission functions for the NICMOS (solid line) and Mauna Kea (dotted line) filter sets.  The curves centered around 1.6 $\mu \rm m$ are the F160W and H filters and the curves centered around 2.2 $\mu \rm m$ are the F222M and K' filters.}
\label{filters}
\end{figure}

We have also made use of archival NICMOS images obtained on 1998 February 02 with the Hubble Space Telescope using the Near Infrared Camera and Multi-Object Spectrometer (NICMOS) as part of the Guaranteed Time Observing (GTO) program 7219 (PI: Scoville; \citealt{sco00}).  The imaging program employed three spectral filters (F110W: $\lambda_{\rm c} = 1.100 \ \mu \rm m$,  FWHM = 0.592 $\mu \rm m$; F160W: $\lambda_{\rm c} = 1.594 \ \mu \rm m$,  FWHM = 0.403 $\mu \rm m$; F222M: $\lambda_{\rm c} = 2.116 \ \mu \rm m$,  FWHM = 0.143 $\mu \rm m$) in NICMOS camera 2 (pixel scale: $x_{\rm scale} = .0762$\arcsec \ pixel$^{-1}$, $y_{\rm scale} = .0755$\arcsec \ pixel$^{-1}$; $x_{\rm scale} =$ 36.3 pc pixel$^{-1}$, $y_{\rm scale} =$ 36.0 pc pixel$^{-1}$).  Four exposures in each filter, using STEP8/NSAMP = 10, 11, and 12 multiaccum sampling \citep{mac97} for F110W, F160W and F222M, respectively, were taken with corresponding integration times of 40s, 48s, and 56s in each exposure.  NICMOS camera 2 critically samples its point spread function at $1.6 \ \mu \rm m$ \citep[see][\S 2.2]{max05}.  All images were acquired under a four-point ``dither" pattern with pointing offsets of 1.9215\arcsec, thus enabling critical sampling at $1.1 \ \mu \rm m$ (after image combination), as well as improved sampling and rejection of data from defective pixels at all wavelengths.  Because the F160W and F222M filters provide wavelength coverage roughly similar to the H and K' Mauna Kea filters (Fig.~\ref{filters}), these two NICMOS images are used primarily as a tool to help perform photometric calibrations on the NIRC2, AO images.  (The details of this photometric calibration technique are described in \S\ref{FluxCalibrations}.)  However where the fields of view of the NICMOS F160W and F222M images extend beyond the NIRC2 fields of view, the NICMOS images provide additional spectral information and thus we analyze these images.  Note that of all the data discussed in this paper the NICMOS images have the lowest spatial resolutions -- about 4 times lower than the resolution provided by the Keck AO system at similar wavebands.  The theoretical FWHM of the diffraction-limited PSFs for the three NICMOS images are 0.093\arcsec, 0.13\arcsec, and 0.18\arcsec, corresponding to linear scales of 44.2, 64.3, and 85.0 pc.

Details regarding both the WFPC2 and NICMOS HST observations are shown in Table~\ref{tableObsHST}.

\section{DATA PROCESSING}\label{Data Processing}
The complete set of data is composed of 7 images from 3 different instruments at the following bandpasses, in order of increasing central wavelength: B, I, F110W, F160W, H, K', and F222M.  The inhomogeneous nature of these images, including the manner in which they were taken, the morphology that they display, and their resolutions which span more than a factor of five, requires that we employ non-uniform processing techniques.  We shall discuss the pitfalls inherent to treating data in a non-uniform fashion.

\subsection{Data Reduction}\label{DataReduction}
The near-infrared images taken with NIRC2 were reduced in the standard way, by masking bad pixels, flat fielding each frame using twilight images, and dark subtracting and sky subtracting frames of the appropriate exposure times.  Before combining NIRC2 narrow camera frames, we removed the geometric distortions using the IRAF DRIZZLE package with the distortion coefficients published in the NIRC2 manual, and the kernel parameter set to ``lanczos3.''

The NICMOS data analyzed in this paper were obtained by recalibrating and reprocessing the archival data.  Instrumentally calibrated count-rate images were created from the raw archival multiaccum frames after applying linearity corrections, dark-frame subtraction, DC bias compensation, flat-field correction, and cosmic-ray rejection.  Darks, made from contemporaneous on-orbit calibration observations, calibration reference files for linearity corrections, obtained from STScI, and flats, obtained from the NICMOS Instrument Definition Team (IDT), were applied in an IDL-based process closely following the IRAF/STSDAS calnica task.  The image data were photometrically calibrated based upon the absolute instrumental calibration established by the IDT (F110W: $2.031 \times 10^{-6}$ Jy count$^{-1}$ s; F160W: $2.190 \times 10^{-6}$ Jy count$^{-1}$ s; F222M: $5.487 \times 10^{-6}$ Jy count$^{-1}$ s).  Known ``bad" or defective pixels were replaced by two-dimensional Gaussian weighted interpolation of good neighbor pixels with wavelength-dependent weighting radii of 3, 5, and 7 pixels for the F110W, F160W, and F222M frames respectively.  The fully calibrated images were geometrically rectified to correct for the $\sim 0.9$\% linear geometrical distortion in NICMOS camera 2, and for each filter, those four images were median combined into a single image on a 2-times finer resampled grid ($\sim 38$ mas pix$^{-1}$).  All image resampling (including re-mapping for registration, and rectification) was done with sinc-function apodized Gaussian weighted interpolation using the IDP3 package developed by the NICMOS IDT \citep{sch02}, conserving the target flux density while resampling.  The three resulting images (one for each filter) are those that are discussed further in this paper.

We obtained the archival WFPC2 data from the WFPC2 associations at MAST (Multimission Archive at STScI), and these data are corrected for bias, dark current, flat fielding, and cosmic rays.  We performed no additional processing on these WFPC2 associations.

\subsection{Cluster Identification}
\begin{figure}[! ht]
\epsscale{1.0}
\plotone{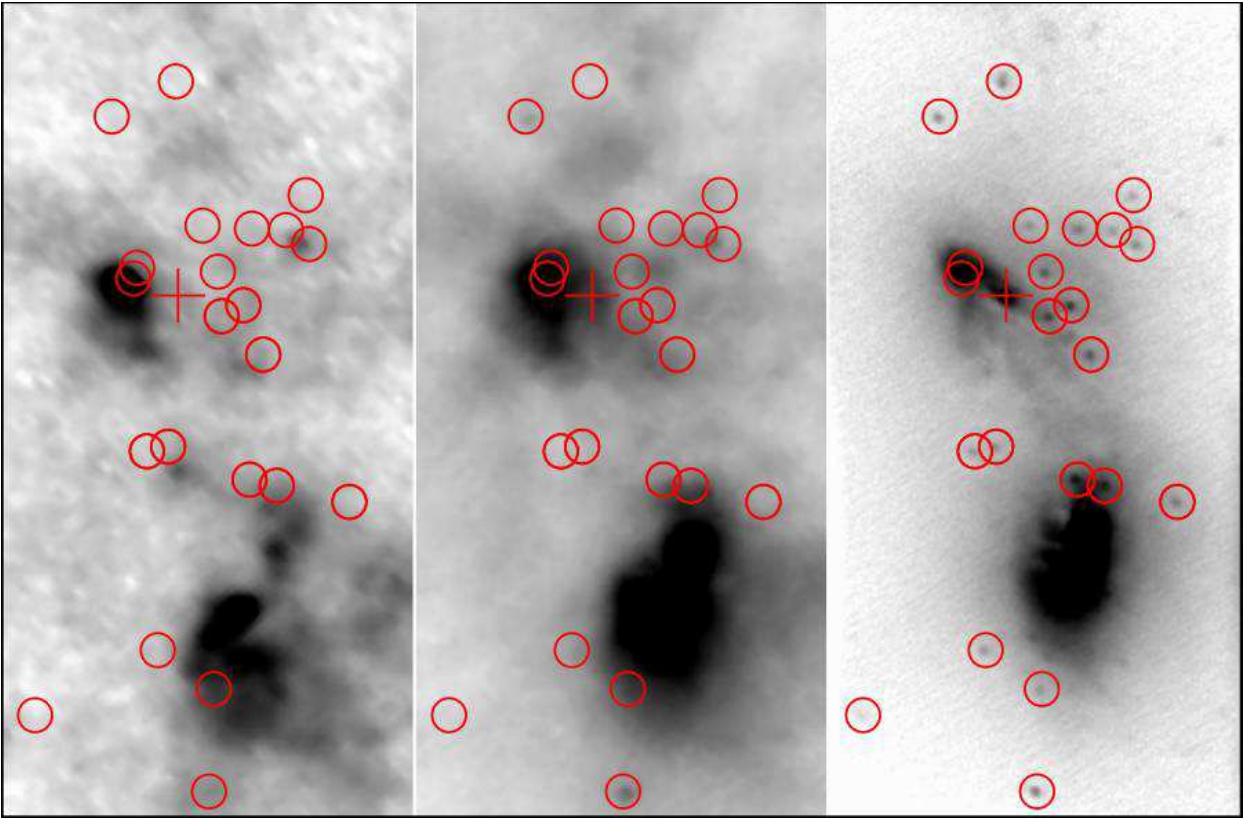}
\caption{Left and middle panels show logarithmically scaled WFPC2 images at B and I bands, respectively.  Known cluster locations are marked with circles.  `$+$' marks the location of cluster \#15 which is coincident with the northern black hole.  Images are 2.5\arcsec $\times$ 5\arcsec \ (1.2 $\times$ 2.4 kpc); North is up and East is to the left.  Right panel shows the same cluster locations superposed on a logarithmically scaled, unsharp masked K' image taken with NIRC2 using Keck NGS AO.  The northernmost and southernmost clusters encircled are clusters \#5 and \#28, respectively.}
\label{grayImage}
\end{figure}

We attempted to identify unresolved star clusters in the 7 images in a systematic way by using the FIND routine adapted for IDL from DAOPHOT.  However, this method and variations on this method, including using the FIND algorithm on unsharp-masked images, were all unsuccessful, especially in regions where the diffuse background light from the galaxy is highly variable or bright.  Therefore in this paper we have visually identified clusters, and we define a cluster to be any source that appears unresolved in an unsharp masked image.  We neglect clusters that are visible only at B and/or I bands because these have been previously analyzed by \citet{pas03}, and our analysis would provide no new spectral coverage.  Figure~\ref{grayImage} shows the cluster positions superimposed on images at B, I, and K' bands.

\begin{figure}[! ht]
\epsscale{1.0}
\plotone{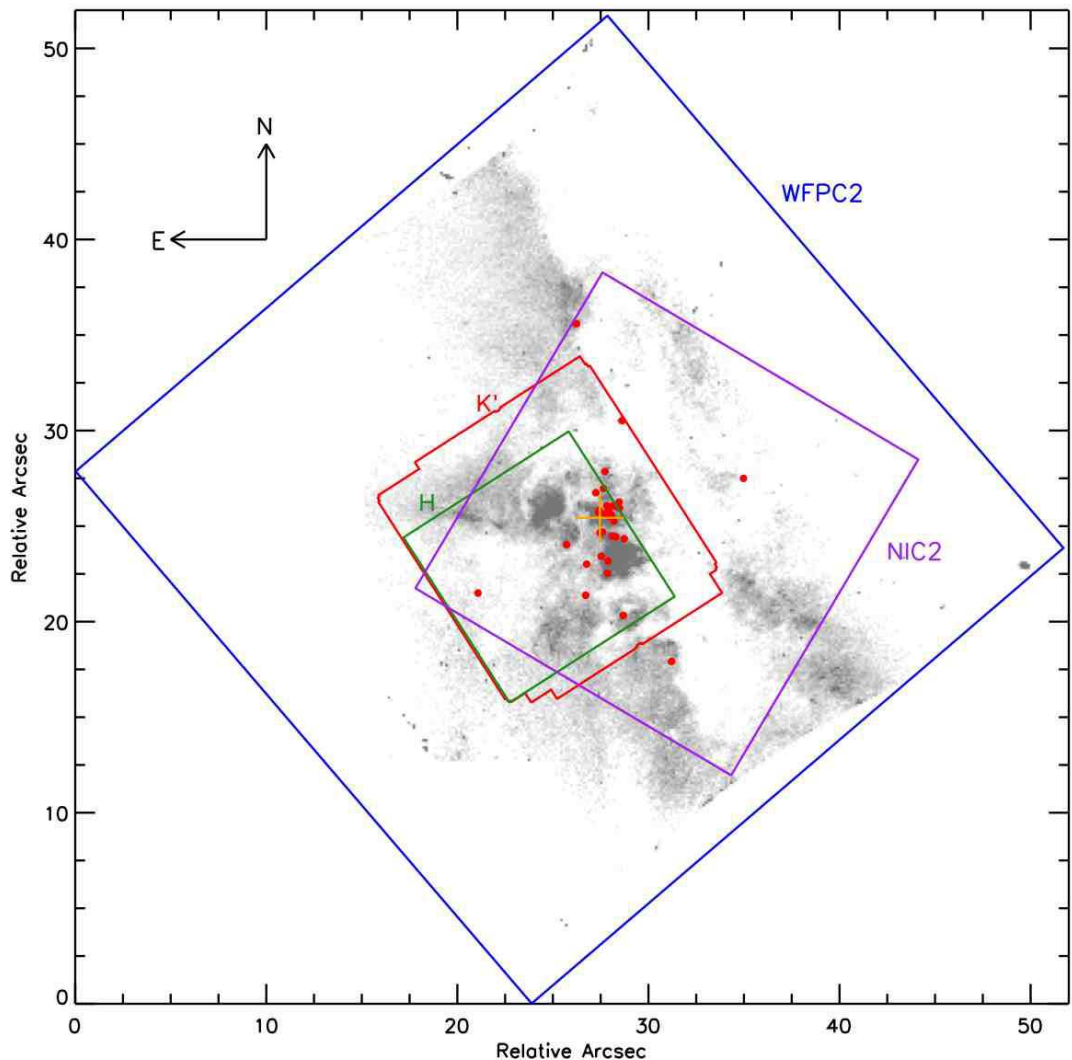}
\caption{Relative fields of view for the complete set of 7 images analyzed in this paper.  The blue square shows the WFPC2 Planetary Camera FOV (F450W and F814W filters).  The purple square shows the NICMOS NIC2 FOV (F110W, F160W, and F222M filters).  The red and green squares show the NIRC2 narrow camera FOV (K' and H filters, respectively).  Red dots show the locations of all 32 clusters analyzed in this paper, for which we have spectral information in at least one of the NIR wavebands.  The orange cross marks the location of the northern black hole, as determined by \citet{max06}.  The overlaid grayscale image shows NGC 6240 taken with the WFPC2 F450W filter.}
\label{FOVfig}
\end{figure}

\begin{figure}[! ht]
\epsscale{1.0}
\plotone{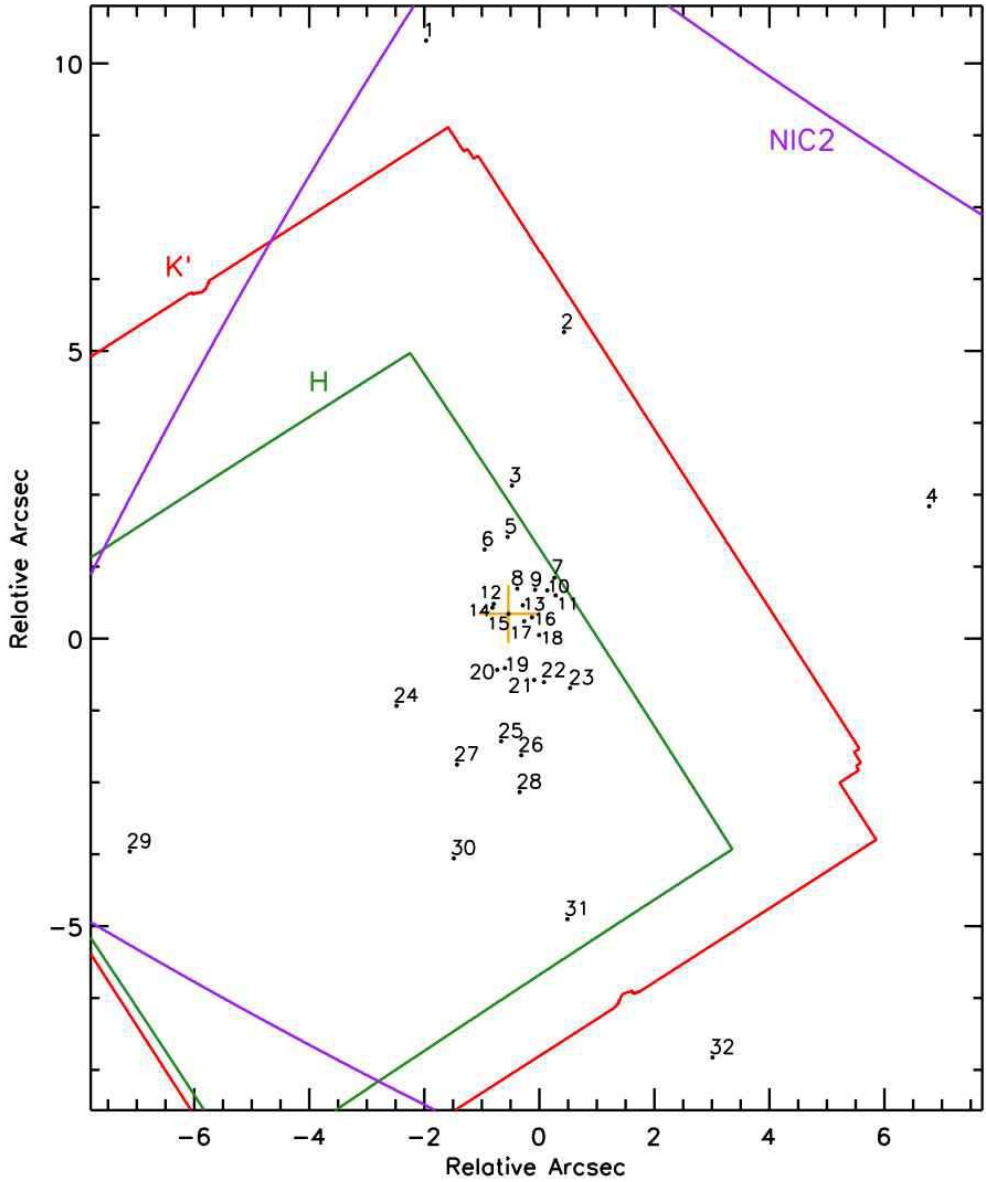}
\caption{Position and identification number of all 32 clusters, shown relative to the infrared fields of view.  Scale in arcseconds is $\sim 1/4$ that in Fig.~\ref{FOVfig}. Identification numbers increase toward the south.  The orange cross indicates the position of the northern black hole which is coincident with cluster \#15.}
\label{ClusterIDs}
\end{figure}

In total we identified 32 clusters that are visible in at least one of the infrared passbands (F110W, F160W, H, K', and F222M).  Of these, 27 lie within the intersection of the 7 fields of view, lending themselves to the most complete spectral analysis.  Of these 27 clusters, only 7 have low enough extinction to be detected at I band, and this number decreases to 5 at B band.  Figure~\ref{FOVfig} shows the locations of all 32 clusters and the relative sizes and orientations of the 7 fields of view.  In Figure~\ref{ClusterIDs} we label each cluster with a number from 1 through 32 with 1 being the northernmost cluster.  Note that cluster \#15 is coincident with the site of the northern black hole, as determined by \citet{max06}.  For reference, in the publicly available WFPC2 I band image, the brightest pixel in the northern nucleus is 0.27\arcsec \ east  and 0.099\arcsec \ north of the northern black hole (i.e. cluster \#15).   In Figures~\ref{FOVfig} and~\ref{ClusterIDs} we mark the position of the northern black hole with an orange cross.  The positions of each cluster relative to cluster \#15 can be found in Table~\ref{clusterTable}.

The complicated morphology of NGC 6240 makes it impossible to specify a single cluster detection limit across the entire FOV.  The morphology of the H and K' images is relatively smooth, with background luminosity increasing sharply toward the double nuclei.  Therefore the cluster detection limits at H and K' worsen monotonically towards the double nuclei.  While it is generally true that the detection limits worsen toward the double nuclei in the B and I images as well, the situation here is more complicated because these images are far more patchy due to dust obscuration.  We estimate our detection limits by adding simulated clusters to various locations in each image, and decreasing the fluxes of the simulated clusters until the clusters become undetectable by eye.  (See \S\ref{FluxCalibrations} Cases I and II for a detailed description of how we create the simulated clusters.)  Using this method, we estimate that clusters brighter than $K_{\rm AB}=22$ mag, $H_{\rm AB}=23$ mag, $I_{\rm AB}=24$ mag, and $B_{\rm AB}=26$ mag should be detectable throughout much of the images.

\subsection{Image Alignment}\label{ImageAlignment}
In order to align all 7 images to the same spatial grid we first visually determined which star clusters are common to pairs of images.  (No stars are visible in the small NIRC2 FOV.)  We fed the centroids of the matched cluster pairs into the IRAF geomap routine, and in this way computed second-order polynomial spatial transformation functions.  We applied the transformations with the IRAF geotran package using bilinear interpolation to rebin the images.  Due to NGC 6240's drastically different morphology between B and K' bands, finding enough clusters in common at these two wavelengths was challenging.  Therefore, instead of aligning the B image directly to the K' image, the B image was first aligned to the I image, and then to the K' image using the transformation computed for the I-K' alignment.  This two step process also ensures that $\rm B - I$ colors are as accurate as possible, with the fewest errors introduced from poor image alignment.  Typical rms fitting errors computed by geomap were $\leq 0.5 \rm \ mas$ when aligning Keck images, $\leq 2.5 \rm \ mas$ when aligning HST images, and $\leq 5 \rm \ mas$ when aligning HST images to Keck images.  

\subsection{Aperture Photometry \& Flux Calibrations}\label{FluxCalibrations}
Because the star clusters, although unresolved, cannot be conveniently modeled as many point sources on a smooth sky background, and because the Strehl ratio for the AO images is uncertain, we performed aperture photometry rather than using the standard PSF-fitting photometry packages like DAOPHOT.  We explored two aperture photometry techniques.  First, we simply performed aperture photometry on all images using an aperture radius of 0.07\arcsec \ and a 0.02\arcsec-wide sky annulus centered at a radius of 0.1\arcsec \  from the cluster centroid.  This annulus is near enough to the cluster that we can accurately measure the local background in a complicated environment, while the aperture is large enough to lessen the effects of poor image alignment.  Second, we tried to minimize the color errors by creating matched-resolution image pairs before performing aperture photometry.  We found that this second technique produced poorer results because we were forced to degrade the image quality of the NIRC2 images, making many of the clusters disappear into the noise.  Therefore, the results presented in this paper were calculated by performing simple aperture photometry on all the clusters, using the same aperture and annulus size for each filter.  Note that the WFPC2 and NIRC2 images have very similar resolutions, with the best resolution (from the H band AO image) being a factor of 2 better than the worst resolution (from the WFPC2 I band image).  However the NICMOS images have poorer resolutions than the WFPC2 and Keck AO images, so that performing aperture photometry on these images with the same size aperture and annulus introduces color errors in the case where different colored background flux from the galaxy falls within the aperture.  Thus caution must be used when interpreting the NICMOS data, especially in regions near the nuclei where the background flux from the galaxy is bright.

In Figures~\ref{seds_7_18} -~\ref{seds_1_32} we present the results of our aperture photometry measurements in the form of SEDs for each cluster, and in the following paragraphs we outline the procedure we used to create these SEDs for two key cases.  Other cases can be deconstructed into variations on these two cases.

\subsubsection*{Case I: Creating an SED for a Cluster Detected Only By HST}\label{CaseI}
In order to populate the SED of a cluster that is detected in only the B, I, F110W, F160W, and F222M images\footnote{A cluster would be detected in only the B, I, F110W, F160W, and F222M images, while not being detected in the H and K' images, if it is located outside of the NIRC2 narrow camera field of view.  Cluster \#4 is an example of such a cluster for which the described procedure was performed.}, first we perform aperture photometry on the cluster in each image (using aperture parameters just described).  We calculate appropriate aperture corrections by repeating this procedure on simulated point sources extracted from the Tinytim (ver 6.3) software package.  Finally, we convert the (aperture corrected) counts to calibrated flux (in Janskys) using the PHOTFLAM and PHOTPLAM keywords in the fits header of the WFPC2 images, and the instrumental calibration values determined by the NICMOS IDT (see \S\ref{DataReduction}).  On an SED graph we represent these measured fluxes in the form of a single point at the position of the central wavelength for each filter.

To calculate the photometry errors shown in Figures~\ref{seds_7_18} -~\ref{seds_1_32}, for each cluster we simulated 100 additional clusters using Tinytim PSFs.  The simulated clusters were made to have the same SED as that which was measured for the real cluster, and they were placed in approximately similar areas of the image as the real cluster.  (We defined 3 regions of the images -- far, close, and intermediate distances from the nuclei.)  Next, we performed the same aperture photometry procedure on the simulated clusters as we did on the real clusters.  The resulting distribution of 100 measurements is sensitive to the exact placement of the simulated clusters and is not gaussian.  Therefore we discard the 10 smallest and largest measured values, and quote errors as 80\% confidence limits from the pared distribution.  The error bars on the single data point at which all models are normalized represent something slightly different from the error bars on the other points in the SED.  The former are 80\% confidence limits on the flux at that frequency, while the latter are 80\% confidence limits on the color.  Thus when analyzing these normalized SEDs, which are plotted in log-linear space, realize that the SEDs may shift vertically by an amount governed by the error bar on the point at which the SEDs are normalized, while the slopes of the SEDs may change only by the size of the error bars on the \emph{other} points.  Since the slope of an SED is governed by the stellar population of that cluster and any intervening dust, the color error bars indicate how confidently we know the cluster's stellar population.  Likewise, since the total mass of the cluster corresponds to an overall multiplicative factor, i.e. a vertical shift in log-linear space, the flux error bars indicate how confidently we know the cluster's mass, assuming no errors on the stellar population and extinction values.

In the error analysis procedure described above, if the 80\% confidence limits indicate that the cluster may have negative flux, as is sometimes the case when performing aperture photometry on barely detected clusters, then we replace the lower error bar with a downward arrow, signifying an upper limit.  (We still plot the upper error bar.)  For clusters that we deem undetected we calculate upper limits using a different method.  We create a simulated cluster using a Tinytim PSF, and place it exactly at the location of the known cluster.  We decrease the flux of the simulated cluster until it is undetectable (by eye) and call the final flux an upper limit.  

\subsubsection*{Case II: Creating an SED for a Cluster Detected Only with Keck AO}\label{caseII}
Because absolute flux calibration of NGS AO images is very difficult when the off-axis Strehl is unknown, the procedure we have adopted for populating a cluster's SED with photometric information taken by the Keck AO system is quite different from the procedure outlined for HST in Case I.  As an example, we outline the procedure used to populate the SED of a cluster detected only in the H and K' images.  Cluster 18 is an example of such a cluster for which the following procedure was employed.


We begin as before by first performing aperture photometry on the cluster in the K' image.  Next, because we have no way of performing an accurate aperture correction or converting counts into Janskys for this image taken using AO, we make use of the NICMOS F222M image.  Though the particular cluster in question is not visible in the lower resolution NICMOS image, one cluster, namely cluster \#2, \emph{is} visible in both images, and is relatively far from the bright nuclei.  We calculate the total flux of cluster \#2 in the F222M image using the standard procedure for NICMOS data, namely performing aperture photometry and applying an aperture correction calculated from a Tinytim PSF, and converting aperture corrected counts into calibrated Janskys by using the appropriate instrumental calibration value determined by the NICMOS IDT.  Next we perform aperture photometry on cluster \#2 in the K' image.  The ratio of the total flux of cluster \#2 to the aperture photometry result can be used to convert from counts in the K' image (calculated using a particular aperture and annulus) to \emph{total}, \emph{calibrated} flux for any cluster.  We use this ratio to convert the counts in the particular aperture and annulus for cluster \#18 in the K' image to \emph{total}, \emph{calibrated} flux.  On an SED graph, this flux is represented in the form of a single point at the position of the central wavelength for the K' filter.

Because the K' and F222M filters are not identical (as can be seen in Fig.~\ref{filters}), the bootstrapping procedure described above produces accurate results only when the slopes of the SED in the NIR are flat, or the same, for clusters 2 and 18.  Indeed, the NIR SED is very flat for the case of young, dusty clusters.  To estimate the errors introduced by this bootstrapping procedure we convolved various \citet{bru03} stellar population models with the transmission functions for the F222M and K' filters.  Assuming a Salpeter IMF, solar metallicity, and Padova evolutionary tracks, we determine that observations taken with the two different filters should differ by less than 0.1 magnitudes for a cluster between the ages of 0 and 10 Gyr.  If an intervening dust screen, corresponding to extinctions ranging from $A_{\rm v}=4$ to 12, reddens the cluster and flattens the NIR SED, then observations taken with the two different filters should differ by less than $K'_{\rm AB}=0.05$ for a cluster that is less than 1 Gyr old.  A completely analogous procedure, which makes use of cluster \#29, is used to bypass the aperture correction and flux calibration of the H band AO images; the errors introduced by calibrating H band data with F160W images are comparable to the 0.1 mag errors quoted above.

The photometry error calculations and the upper limit calculations in the case where the cluster is only detected in the AO images are completely analogous to the procedures described in Case I.  However, instead of using Tinytim PSFs to simulate clusters, we used a theoretical AO PSF.  We model the AO PSFs as the sum of two gaussians.  One gaussian has a FWHM equal to $0.98\lambda/D$ where $\lambda$ is the observing wavelength and D is the 10m diameter of the Keck telescope.  The other gaussian has FWHM equal to $0.98\lambda/r_{0}$, where $r_{0}$ is the Fried parameter, which we assume to be 30 cm.  We let the on-axis Strehl ratio govern the ratio of the total amounts of energy in each gaussian.  We recognize that the Strehl used for the simulated clusters is certainly over estimated.  One of the consequences of these assumptions is evident in Figure~\ref{seds_1_32}, where upper limits for cluster \#31 are clearly far too low.  (Larger Strehls make fainter clusters easier to detect.)  
Therefore when interpreting infrared upper limits shown in Figures~\ref{seds_7_18} -~\ref{seds_1_32}, the upper limits computed from NICMOS data -- for which the PSF is well modeled by Tinytim -- should be trusted more than the upper limits computed from the AO data.

\section{RESULTS}\label{Results}
In this complicated galaxy merger the values we have calculated, namely a cluster's magnitude in \emph{at most} 5 distinct wavebands,\footnote{Recall that the F160W and H wavebands are similar, as are the F222M and K' wavebands.} are outnumbered by the unknown variables, which include, but are not limited to, the following list: 1) the metallicity of gas that formed each cluster; 2) the duration and temporal profile of the starburst for each cluster; 3) the initial mass function (IMF) of stars in each cluster; 4) the stellar evolutionary track followed by every star in each cluster; 5) the extinction, scattering, and absorption properties of the dust in and in front of each cluster; and 6) the mass and age of each cluster.  In principle we would like to find
 the stellar population that most closely matches the observed photometry for each cluster, but because it is impossible to fully constrain this problem, rather than perform minimization techniques on a many-dimensional parameter space we adopt a different approach which requires making the following simplifying assumptions.  We assume that all clusters: 1) have solar metallicity; 2) were formed in an instantaneous burst; 3) are composed of stars with stellar masses governed by a Salpeter IMF; 4) are composed of stars that evolve on Padova (1994) evolutionary tracks; and 5) have evacuated all intracluster dust before an age of $\sim$3 Myr\footnote{Note that even if some intracluster dust remains after 3 Myr, our assumption is still good assuming that the effect of the remaining intracluster dust is far less than that of the dust along the line of sight through the merging galaxy.}, and are obscured behind a simple dust screen that behaves according to $R_{V}=3.1$.

The fact that all 32 of these clusters are concentrated within $<5$ kpc of the double nuclei (in projection) suggests that their formation was triggered by the merger and that they were formed roughly contemporaneously.  To simplify our analysis we initially assume that all clusters were formed simultaneously, although this is obviously false.  Later we relax this assumption to allow for clusters of three possible young (or intermediate) ages.  It is possible that some subset of the clusters are old GCs from the progenitor galaxies that happen to be passing through the center of the merger, but this is surely not the case for the majority of the clusters -- especially those closest to the nuclei. Assuming the clusters are not old GCs, it is reasonable to choose a metallicity close to solar.  If the triggering event was short, then the assumption that the clusters' stellar populations are well modeled by an instantaneous burst is roughly accurate.  However, those clusters that are located very near to the double nuclei may be experiencing a continual accretion of gas, and therefore it is possible that they would be better modeled by a stellar population with a constant SFR.  We leave the clusters' ages and the amount of intervening dust along the line of sight to each cluster as free variables.

\begin{figure}[! ht]
\epsscale{1.0}
\plotone{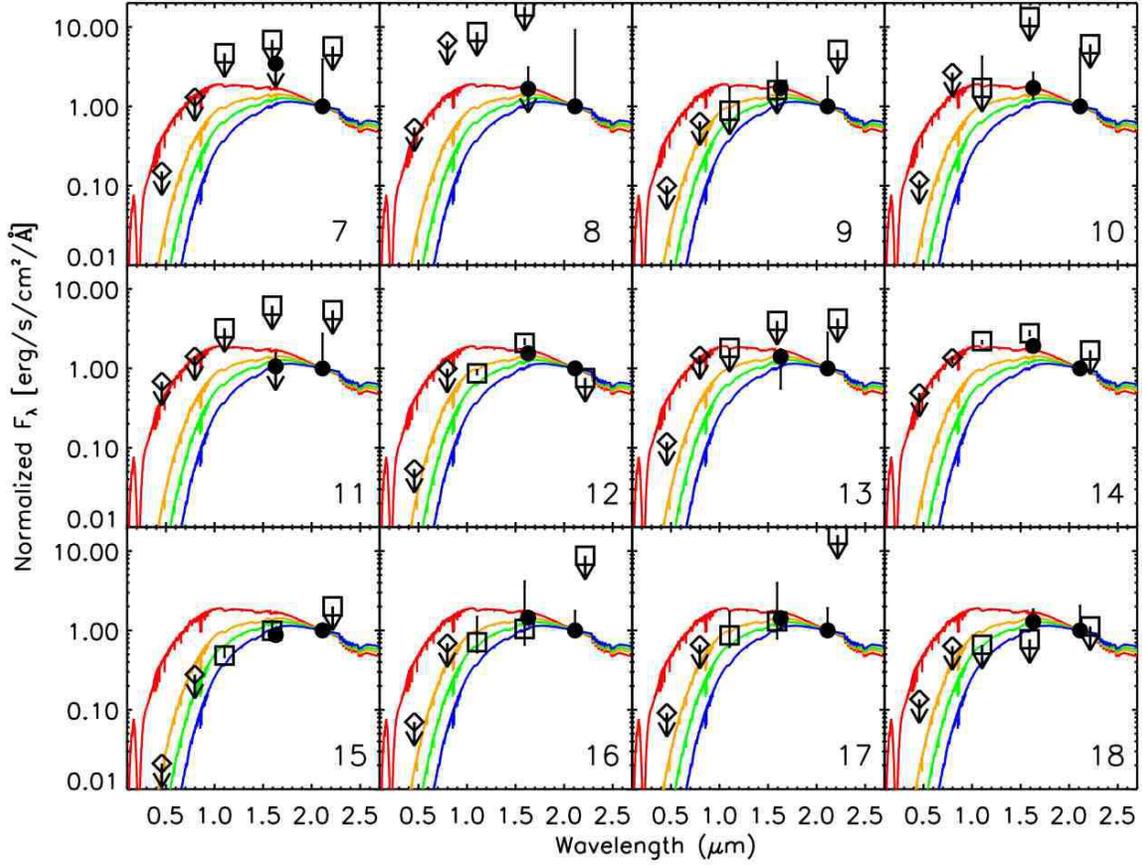}
\caption{Open diamonds, open squares, and filled circles mark the photometric measurements made for 12 clusters using WFPC2, NICMOS, and NIRC2 images, respectively.  The 12 clusters depicted here are those situated nearest to the northern nucleus, in projection; cluster labels are given in the lower right corner of each panel.  Error bars show 80\% confidence limits.  Upper limits are depicted with a downward arrow for those filters where a cluster is not detected.  Read \S\ref{FluxCalibrations} to understand why some clusters show both upper error bars and upper limits for a single filter.  Over-plotted on the measured data points are model SEDs created by the population synthesis code of \citet{bru03}.  The model SEDs show 14.5 Myr-old instantaneous burst, solar metallicity, stellar populations, assuming a Salpeter IMF and Padova (1994) evolutionary tracks.  Models have been extinguished by varying column densities of dust in a simple dust screen, following the \citet{dra03} dust prescription.  Red, orange, green, and blue curves correspond to $\rm A_{V}=3.5$, 6.5, 8.5, and 11 mag, respectively.  All model SEDs, as well as the observed K' photometry, are normalized to the K' data point for ease of comparison.}
\label{seds_7_18}
\end{figure}

\begin{figure}[! ht]
\epsscale{1.0}
\plotone{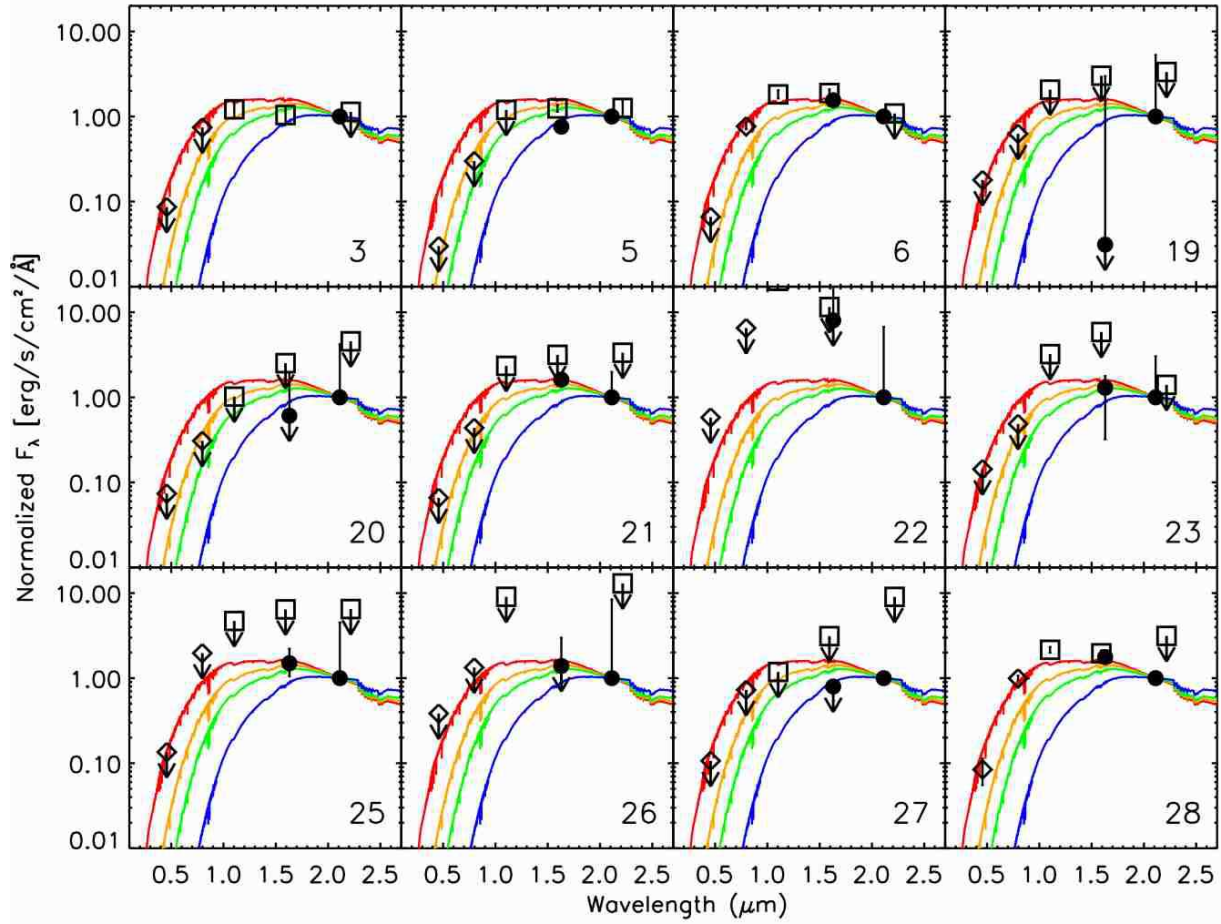}
\caption{Similar to Figure~\ref{seds_7_18}, but here we show the 12 \emph{next} closest clusters to the northern nucleus.  Red, orange, green, and blue curves correspond to 14.5 Myr old stellar populations with $\rm A_{V}=4.5$, 6.5, 8.5, and 14 mag, respectively.}
\label{seds_3_28}
\end{figure}

\begin{figure}[! ht]
\epsscale{1.0}
\plotone{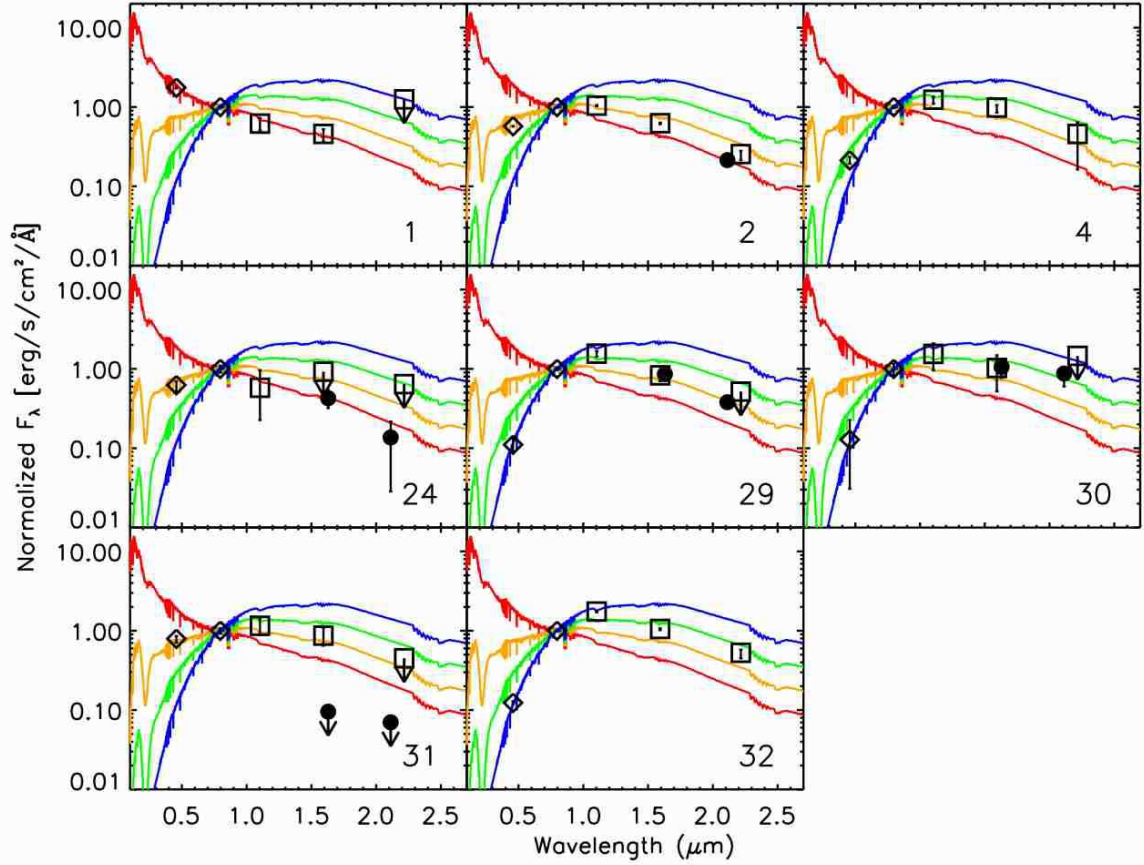}
\caption{Similar to Figure~\ref{seds_7_18}, but here we show the 8 clusters farthest from the northern nucleus, with all photometry and model SEDs normalized to the I data point for ease of comparison.  Red, orange, green, and blue curves correspond to 14.5 Myr old stellar populations with $\rm A_{V}=0.5$, 2.0, 3.5, and 5 mag, respectively.  Read \S\ref{FluxCalibrations} (case II) to understand why the H and K' upper limits shown for cluster \#31 should not be trusted.}
\label{seds_1_32}
\end{figure}

After making the aforementioned simplifications, our approach entails visually comparing the SEDs that resulted from performing aperture photometry at B through K' bands, to model SEDs created by the isochrone synthesis code of \citet{bru03}.  In Figures~\ref{seds_7_18} -~\ref{seds_1_32} we present the results of this approach for the case where we assume that all clusters are 14.5 Myr old.  (We justify this choice of age shortly.) We organize the results into 3 groups; Figure~\ref{seds_7_18} shows the 12 clusters closest to the northern nucleus, Figure~\ref{seds_3_28} shows the 12 next closest clusters, that lie at an intermediate distance from the norther nucleus, and Figure~\ref{seds_1_32} shows the remaining 8 clusters that are farthest from the northern nucleus.  All model SEDs shown in these figures embody the former 6 simplifying assumptions, and for the purpose of visual comparison, each SED has been scaled so that it passes through either the K' or the I data points.\footnote{All models are for zero redshift.  The small redshift of NGC 6240 $(z=.024)$ implies an insignificant k-correction.}  The different colors of the SED curves correspond to varying column densities of interstellar dust applied using the \citet{dra03} dust prescription with $\rm R_{V} = 3.1$.  (Grain size distribution comes from \citet{wei01}; PAH optical properties come from \citet{li01}.)  For the 12 clusters nearest the northern nucleus, column densities corresponding to between 3.5 and 11 magnitudes of visual extinction were required to match the observations with the models.  Similarly, when comparing the intermediate-distance clusters to the model SEDs, column densities corresponding to $\rm A_{V}$ between 4.5 and 14 mag were applied to the models.  But for the clusters farthest from the northern nucleus, much smaller column densities, corresponding to $\rm A_{V}=0.5-5$ mag, were needed to produce SEDs spanning the range of observations.  Immediately we see that if our simplifying assumptions are correct, and if all clusters are assumed for the time being to be the same age of 14.5 Myr, then on average the outermost clusters are obscured by a thinner dust screen than that which obscures the inner clusters.  

We repeat this procedure for two additional cases, where all clusters are 2.88 Myr old, and where all clusters are 180 Myr old.  These ages were chosen for the following two reasons. 1) The ages span a range that is consistent with our previous assumptions.  For example, clusters younger than 3 Myr could have considerable intracluster dust, and clusters older than several hundred million years would have been formed before the onset of the merger, thus it is unlikely that they would have formed contemporaneously or be situated coincidentally near the nuclei\footnote{Note that in principle the clusters could have been formed during an initial tidal interaction just a few hundred million years ago, making them now older than 180 Myr.  However, as will be discussed in \S\ref{Discussion}, the masses required to match the observations to the models for clusters older than $\sim$180 Myr are too large to be credible.}.  2) The ages bracket the results published in the literature for other star clusters in NGC 6240.  Using photometry at the F450W, F547W, and F814W wavebands, \citet{pas03} found that the clusters in NGC 6240's main body and tidal tails have probable ages ranging from 5 to 13 Myr.  By analyzing the CO 2-0 absorption bandhead equivalent width, \citet{tec00} show that NGC 6240's K-band light is dominated by a $\sim$20 Myr-old, short-duration starburst.

For each of the three ages analyzed, we choose as the ``best-fitting" model the SED with the smallest value of $\rm A_{V}$ that matches the observations for each cluster.  For example, the observed SED for cluster \#13 in Figure~\ref{seds_7_18} is consistent with model SEDs having $\rm A_{V}$ ranging from 5.5 to 16 mag; therefore we assume that cluster \#13, \emph{if it is 15 Myr old}, is extinguished by $\rm A_{V}=5.5$ mag.  Note that larger extinctions translate to more massive clusters, so our method finds the least massive cluster that matches the data for a given age.

\begin{figure}[! ht]
\epsscale{.7}
\plotone{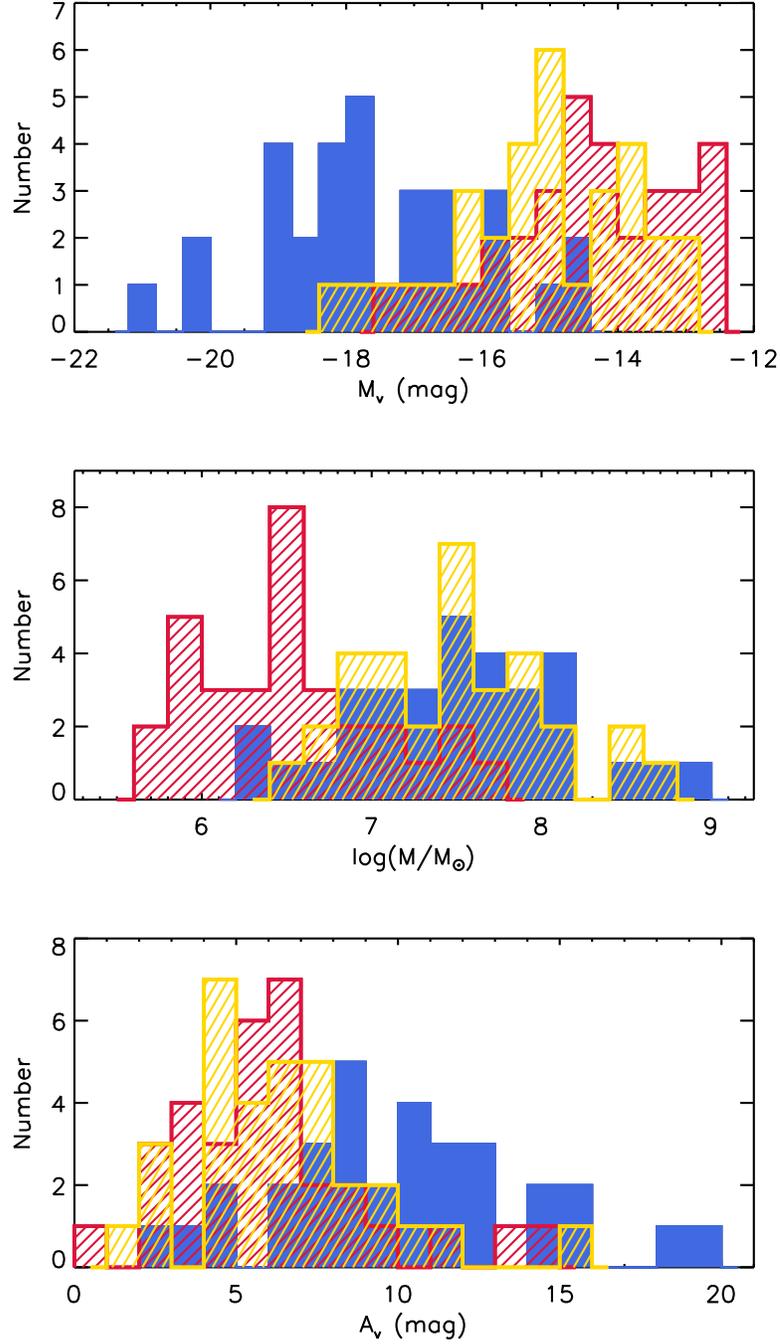}
\caption{Red histograms show the un-extincted absolute V-band magnitudes, masses, and extinctions of clusters when all clusters are forced to be 14.5 Myr-old, and differ only by the amount of extinction applied to each.  (e.g. Red histograms summarize the results of Figures~\ref{seds_7_18}-\ref{seds_1_32} for the best-fitting $\rm A_{V}$.)  Blue and yellow histograms show similar summaries for the cases where we force all clusters to be either 2.88 Myr or 180.5 Myr old, respectively.  Note that these histograms include data for all 32 detected clusters, although we believe 5 of the clusters may be better described by different, intermediate ages.}
\label{histograms}
\end{figure}

We do not show graphs of the 2.88 and 180 Myr-old model SEDs for every cluster; instead we tabulate the results of this exercise in Table~\ref{clusterTable}.  The values of $\rm A_{V}$ listed for each cluster and for each assumed cluster age correspond to the ``best-fitting" model just described.  The masses of each cluster are directly proportional to their un-extincted fluxes; for two clusters with equal ages and equal apparent magnitudes, the more extinguished cluster is more massive.  In Figure~\ref{histograms} we depict these same results in the form of histograms.  The blue, red, and yellow histograms show the ``best-fitting" $\rm A_{V}$, and corresponding mass and (un-extinguished) $\rm M_{V}$, of all 32 clusters for the cases where all clusters are assumed to be 2.88, 14.5, and 180 Myr old, respectively.

\begin{figure}[! ht]
\epsscale{1.0}
\plotone{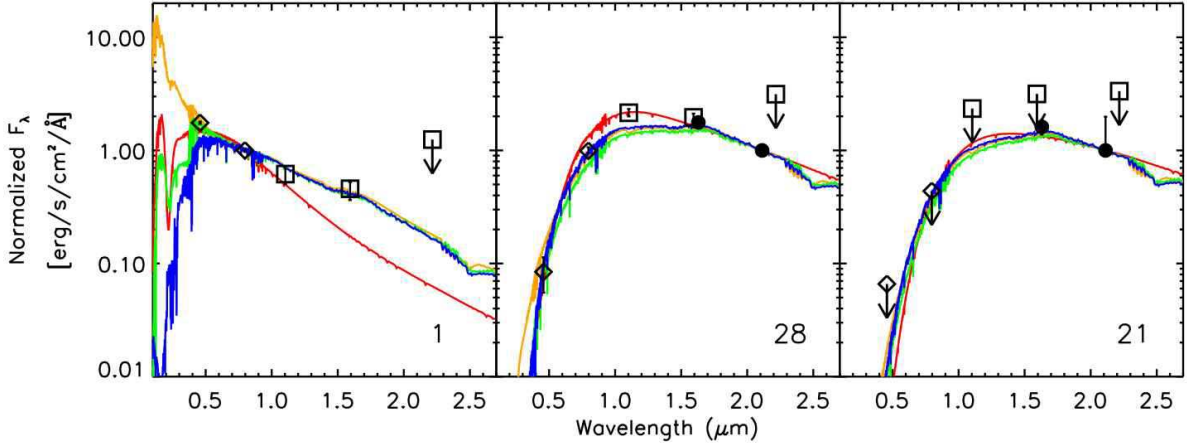}
\caption{Open diamonds, open squares, and filled circles mark the photometric measurements made for clusters \#1, 28, and 21 using WFPC2, NICMOS, and NIRC2 images, respectively.  Over-plotted SEDs in red, orange, green, and blue show 2.88, 14.5, 180, and 2000 Myr-old instantaneous burst stellar populations, respectively.  For each model SED plotted, we have applied the amount of extinction necessary to make the SEDs most closely match the observed photometry.  For cluster \#1, the red, orange, green, and blue curves correspond to $\rm A_{V}=2.5$, 0.5, 1, and 0 mag.  For cluster \#28, the red, orange, green, and blue curves correspond to $\rm A_{V}=8.0$, 4.5, 5.5, and 4 mag.  For cluster \#21, the red, orange, green, and blue curves correspond to $\rm A_{V}=11$, 6.5, 7.5, and 6 mag.  Read Figure~\ref{seds_7_18} caption for more details.}
\label{examplaryPlots}
\end{figure}

While the values listed in Table~\ref{clusterTable} indicate the stellar population with the ``best-fitting"  $\rm A_{V}$ assuming 3 different cluster ages, we have not yet specified which age stellar population best matches the observations.  We address this next step using Figure~\ref{examplaryPlots}, in which we use 3 typical clusters to directly compare the model SEDs with the best-fitting $\rm A_{V}$ for 4 different ages.  Each of the measured SEDs for clusters \#1, 28, and 21 are shown over-plotted with model SEDs corresponding to stellar populations that are 2.88, 14.5, 180, and 2000 Myr old.  Cluster \#1 is typical of a cluster for which it is difficult to distinguish between the 14.5 and 180 Myr-old populations, but it is easy to rule out the youngest and oldest ages with the help of spectral information at wavelengths beyond 1$\mu \rm m$.  In cases like these, to be consistent with our previous method of quoting the least massive clusters that match the data, we choose as the most likely cluster age that which yields the smallest cluster mass.  For cluster \#1, the 14.5 and 180 Myr-old populations correspond to $\sim1 \times 10^{6} \rm M_{\odot}$ and $8 \times 10^{6} \rm M_{\odot}$, respectively, and so we designate this cluster as having aged 14.5 Myr.  In Table~\ref{clusterTable} we indicate this final designation with bold text.

The center panel in Figure~\ref{examplaryPlots} uses cluster \#28 to demonstrate the case where the 2.88 Myr-old model SED visibly fits the observations better than stellar populations of the other three ages.  In cases like these, although the cluster's young age requires it to be very massive, we nevertheless designate it as having aged 2.88 Myr by using bold text in Table~\ref{clusterTable}.  Finally, cluster \#21 is representative of the majority of the clusters, for which it is not possible to rule out any age stellar population through visual inspection of the various model SEDs.  In cases like these, we again resort to choosing the cluster age which yields the smallest cluster mass.  For cluster \#21, the 2.88, 14.5, 180, and 2000 Myr-old populations correspond to clusters that have approximate masses of $70 \times 10^{6} \rm M_{\odot}$, $6 \times 10^{6} \rm M_{\odot}$, $60 \times 10^{6} \rm M_{\odot}$, and $150 \times 10^{6} \rm M_{\odot}$, respectively, and therefore we designate this cluster (and clusters like this one) as having aged 14.5 Myr.  Note that in order for the younger models to be consistent with the observations we must assume more intervening dust; the 2.88 Myr-old model shown for cluster \#21 corresponds to $\rm A_{V}=11$ mag, while the 2000 Myr-old model corresponds to $\rm A_{V}=6$ mag.  This, of course, demonstrates the typical astronomer's dilemma of distinguishing between young, dusty populations and older stellar populations.

Finally, having just described the method we used to estimate the properties of each cluster, we reiterate our motivation to \emph{not} perform a detailed $\chi^{2}$ minimization when matching model SEDs to our observations.  Given the number of assumptions we are required to make in this work, a detailed minimization of the SED fits would provide very little additional insight, and could be misleading.  In particular, a minimization would only describe which of the three chosen ages best fits the models; a fourth (intermediate) age could fit the models better than any of the three.  Therefore it is more instructive to examine the cases of young and old clusters and to discuss how our results would differ if we had chosen different ages.   We do this in \S\ref{ClusterProperties}, and indeed, we discover -- not unexpectedly -- that a different intermediate age may best describe some of the clusters.

\section{DISCUSSION}\label{Discussion}
\subsection{Properties of the Cluster Population}\label{ClusterProperties}

As shown with bold text in Table~\ref{clusterTable}, the best-fitting model stellar populations and dust screens imply that the clusters are consistent with being young ($\sim 3 - 15$ Myr) and having masses that range from approximately $7 \times 10^{5} \rm M_{\odot}$ to $1 \times 10^{8} \rm M_{\odot}$.  The summed mass of all 32 clusters is $\sim4 \times 10^{8} \rm M_{\odot}$, with the single most massive cluster making up $\sim30$\% of the total mass.  The 8 next most massive clusters with $(1 < \rm M < 6) \times 10^{7} \rm M_{\odot}$ contribute $\sim60$\% of the total mass.  Cluster \#15, which is coincident with the site of the northern black hole, is the only cluster which we expect may have a non-stellar SED, or may be poorly modeled by an instantaneous burst.  This cluster is the second most massive cluster and accounts for approximately 10\% of the total mass.

With the exception of the most massive cluster, the clusters analyzed in this paper have masses similar to, albeit slightly larger than, those found in other well-studied merging galaxies.  \citet{whi02} estimate that the young, optically bright clusters in the Antennae galaxies have masses ranging from $2 \times 10^{4} \rm M_{\odot}$ to $4 \times 10^{6} \rm M_{\odot}$.  \citet{gil00} find that the brightest near-IR star cluster in the Antennae is $\sim 4$ Myr old and has $\rm M=1.6 \times 10^{7} \rm M_{\odot}$.  Similarly, many of the young ($<10$ Myr) clusters in Arp 220 have masses $>10^{6} \rm M_{\odot}$, or even as much as $10^{7} \rm M_{\odot}$ \citep{wil06}.  (Note that all three of these aforementioned works assume a Salpeter IMF.)  \citet{mcc03} found kinematic masses of $3.5 \times 10^{5} \rm M_{\odot}$ and $1.5 \times 10^{6} \rm M_{\odot}$ for two clusters in M82.

Recall that the masses stated in this paper are underestimated in the sense that we adopted stellar population models with the smallest value of $\rm A_{V}$ possible.  In addition, for all cases where the observed SEDs fit the different aged models equally well, we designated the correct model to be that which yielded the least massive cluster.  But by comparing the resulting cluster masses to those in the literature, we judge that these two assumptions were probably necessary for the majority of the clusters.  Figure~\ref{histograms} shows that if we had assumed the clusters were much younger or older than $\sim15$ Myr, many of the resulting masses would be far larger than those measured in any other merging system.  In fact, the clusters' masses put stringent constraints on their maximum ages, assuming we have correctly modeled other properties of the stellar populations.  As clusters age past 180 Myr, they become dimmer and dimmer so that the masses required for us to detect them become unreasonably large ($\sim 10^{9} \rm M_{\odot}$).  

The star formation rates measured for NGC 6240 using a variety of methods put additional stringent constraints on the cluster's ages.  \citet{bes01} use the 1.4-GHz luminosity to show that $\rm SFR(M \geq 1 M_{\odot}) = 83.1 \ M_{\odot} \  yr^{-1}$ in the nuclear region of NGC 6240.  \citet{gao04} show that the HCN luminosity, which traces dense molecular gas, is a good indicator of the star formation rate in spirals and ULIRGs alike.  Using the relationship described by \citet{gao04} and the value of $\rm L_{HCN}$ measured by \citet{sol92}, we calculate that the total star formation rate in NGC 6240 is approximately $200 \ \rm M_{\odot} \  yr^{-1}$.  This value is slightly larger than that derived from NGC 6240's integrated IR luminosity, $\sim 140 \ \rm M_{\odot} \ yr^{-1}$ \citep{hec90}.  

The total star formation rate due to the 32 clusters analyzed in this paper is $\sim90 \ \rm M_{\odot} \  yr^{-1}$.  Considering that our measurement exceeds the SFR measured by \citet{bes01} for the nuclear region of NGC 6240, and that the unresolved clusters do not include the majority of the infrared nuclear light (see Fig.~\ref{grayImage}), our measured SFR may be implausibly high.  However, $\sim 80$\% of the total SFR from the clusters comes from just 4 clusters which each contribute SFR $> 6 \ \rm M_{\odot} \  yr^{-1}$.  These 4 clusters are the only clusters for which we determined that the 3 Myr-old model SEDs fit the observations better than models of other ages.\footnote{All four of the of the clusters that have SFR $> 6 \ \rm M_{\odot} \  yr^{-1}$ also have $\rm M > 10^{7} \rm M_{\odot}$.}  We have already conceded that our method of visual inspection does not search the entire parameter space for a best fitting model.  While a Salpeter IMF or a solar metallicity may be to blame, it is unlikely that all but 4 clusters can be reasonably modeled by our assumptions.  Thus we suggest that these 4 clusters are neither 3, nor 15, nor 180 Myr old, but rather some other intermediate age.  The K band brightness implied by the instantaneous burst stellar population models \citep{bru03} monotonically decreases for population ages greater than $\sim10$ Myr, and increases for population ages between $\sim3$ and 10 Myr.  Thus, these 4 clusters may be better described by a stellar population with an age of $\sim10$ Myr, implying that they are less massive and contribute less to the total SFR.

The total SFR measured for the remaining 28 clusters, for which the observations are consistent with 15 Myr-old stellar populations, is $\sim 15 \rm M_{\odot} \ yr^{-1}$.  This agrees well with the literature, and implies that the newly detected clusters account for $\sim 20$\% of the nuclear SFR, or $\sim 10$\% of the total SFR in the ULIRG.  Finally, if we assumed that the same 28 clusters were all 3 Myr old and summed their appropriate star formation rates in Table~\ref{clusterTable}, we would conclude that they contribute a whopping $\sim879 \rm M_{\odot} \ yr^{-1}$.  Thus given our assumptions (outlined in detail in paragraph 1 in \S\ref{Results}) the total star formation rate for NGC 6240 puts rather stringent constraints on the clusters' ages.

To summarize, the masses of star clusters observed in other mergers like Arp 220 and the Antennae strongly suggest that the clusters in NGC 6240 are younger than 180 Myr, while the SFRs measured for NGC 6240 using $\rm L_{\rm 1.4 \ GHz}$, $\rm L_{HCN}$, and $\rm L_{8-1000 \mu m}$ require that the majority of the clusters be older than 3 Myr.  
28 of the 32 detected clusters are consistent with being 15 Myr old, while the large SFRs of 4 clusters (clusters \#6, 28, 29, and 32\footnote{Note that because clusters \#29 and 32 lie outside of the ``nuclear" region for which \citet{bes01} measured $\rm L_{\rm 1.4 \ GHz}$, their high SFRs do not directly contradict $83.1 \ M_{\odot} \  yr^{-1}$ limit set by \citet{bes01}.}) suggest that these 4 may be more consistent with another intermediate age.  The large mass of cluster \#15 and the fact that cluster \#15 is coincident with the northern black hole suggests that the measured SED may have a non-stellar component.  NGC 6240's cluster population -- excluding the 5 aforementioned clusters -- is very similar to other cluster populations in the literature.  The cluster masses range from approximately $7 \times 10^{5} \rm M_{\odot}$ to $4 \times 10^{7} \rm M_{\odot}$, and the total contribution to the SFR from these 27 clusters is $\sim10 \rm M_{\odot} \ yr^{-1}$, or $\sim10$\% of the total SFR in the nuclear region.

The cluster population discussed in this paper is considerably more massive than that which was discussed in \citet{pas03}.  The latter consists of 41 clusters located in NGC 6240's ``main body" and 13 clusters located in the galactic tails.  \citet{pas03} estimate that these clusters have probable masses of $(1-2) \times 10^{5} \ \rm M_{\odot}$.  Because \citet{pas03} do not give the locations of the clusters they analyze\footnote{\citet{pas03} did not publish the locations of the clusters they analyze, nor did they provide this information on the website they mention.  We were not able to obtain the cluster positions from private communication with the authors.} it is difficult to know whether this is a discrepancy or whether we are simply noticing an increase in the cluster mass toward the double nuclei, as was noted by \citet{pas03}.  While it is obvious that \citet{pas03} did not analyze the majority of the clusters discussed in this paper, for which we measure only upper limits on the B and I photometry, it is intriguiging that clusters \#2 and 4, which are located in what appears to be the ``main body" of NGC 6240, and for which B and I measurements were made, are rather massive $(>10^{6} \ \rm M_{\odot})$ compared to estimates made by \citet{pas03}.  It is possible that this apparent discrepancy is due to the fact that \citet{pas03} used Starburst99 stellar population models with a different reddening law and sub-solar metallicity.  This apparent discrepancy cannot, however, be attributed to a difference in IMF, since \citet{pas03} used the same IMF as we did in this paper, namely a Salpeter IMF with lower and upper mass cutoffs of 0.1 and 100 M$_{\odot}$, respecitvely.  

Finally, we note that while using a Salpeter IMF is helpful when comparing our results to the literature, the \citet{kro01} IMF is more realistic and results in lower cluster mass estimates.  Since the infrared luminosity of clusters is dominated by light from high mass stars that quickly turn off the main sequence, and since the Salpeter and \citet{kro01} IMFs differ only for masses below 0.5 M$_{\odot}$, the cluster masses one derives when employing these IMFs differ by a simple scaling factor for clusters with ages less than the main sequence lifetime of a 0.5 M$_{\odot}$ star.  The exact mass scaling can be easily derived by piecewise-integrating the two IMFs from 0.1 to 100 M$_{\odot}$.  The result is that two young clusters with equal infrared luminosities but different IMFs have masses that differ by a factor of $\sim 1.3$, with the larger mass corresponding to the cluster formed with a Salpeter IMF and the smaller mass corresponding to the cluster formed with a \citet{kro01} IMF.  Thus if we had used a more realistic IMF throughout our analysis in this paper, we would have estimated cluster masses to be smaller by a factor of about 0.77.  Correspondingly, our SFR estimates would decrease by this same factor.

\subsection{Dust and Extinction in NGC 6240}\label{DustDiscussion}

NGC 6240's cluster population, if in fact obscured by a simple dust screen with $\rm R_{V}=3.1$ as described by \citet{dra03}, is extinguished by column densities of dust corresponding to $\rm A_{V}$ between 0.5 and 14 mag.\footnote{We have not accounted for Galactic extinction when estimating these values.  The Galactic extinction toward NGC 6240 is $\rm A_{V}=0.25$ mag and is negligible given the uncertainty on our measurements.}  There is no significant difference in the values of $\rm A_{V}$ measured for the 24 clusters closest to the northern nucleus, but the 8 clusters farthest from the nucleus exhibit significantly less extinction.  This is to be expected for an active merger, in which gas is funneled onto the nuclei, in this case powering the AGN that \citet{kom03} detected.  Excluding clusters \#6, 28, 29, 32, and 15 as previously discussed, on average the 6 clusters farthest from the northern nucleus are extinguished by $\rm A_{V}=2.3$ mag, and the remaining 21 clusters closest to the nucleus are extinguished by $\rm A_{V}=7$ mag.

\citet{ger04} and \citet{tec00} both estimate extinctions toward the northern and southern nuclei assuming foreground dust screens.  For the northern nucleus we can directly compare our results to theirs using clusters \#12 and 14 as two sight lines through this region.  (\citeauthor{ger04} and \citeauthor{tec00} both classify the brightest northern knot as the northern nucleus.) \citeauthor{ger04} and \citeauthor{tec00} estimate that $\rm A_{V}=2.35$ mag and $\rm A_{V}=1.6$ mag, respectively.  Both of these values are considerably less than the measurements made in this paper: $\rm A_{V}=6.5$ mag and $\rm A_{V}=3.5$ mag toward clusters \#12 and 14, respectively.  In addition, \citeauthor{tec00} measure a peak extinction of $\rm A_{V}=7.2$ mag, while we find that many clusters discussed in this paper have $\rm A_{V}>7$ mag.  For these reasons we suggest that \citeauthor{ger04} and \citeauthor{tec00} have both underestimated the amount of extinction in this ULIRG.  Our values of extinction agree better with the values in the literature, measured using a variety of methods.  \citet{vig99} used X-ray observations and found $\rm A_{V} \sim 10$ mag, \citet{gen98} used line ratios and found $\rm A_{V} \geq 5$ mag, and \citet{rie85} used continuum K and L spectrophotometry and inferred $\rm A_{V} \sim 15$ mag.  The cause of the previous underestimates may be due in part to the different spatial scales being probed, and in part to the fact that the measurements can only probe the optically thin layer surrounding the nuclei.  The observations presented in this paper are the first infrared observations to resolve the northern nucleus, and thus it is sensible that we would infer higher values of extinction than were found by \citet{tec00}.

Finally, note that while \citet{tec00} may have underestimated the overall extinction in this ULIRG, the morphology of their extinction map is consistent with our findings.  They find that the extinction is greatest between the nuclei and least toward the northern nucleus.  We detect a cluster coincident with the northern nucleus and we find a large concentration of clusters just northwest of the northern nucleus, yet we cannot resolve the southern nucleus, nor do we detect many clusters directly between the nuclei.  Thus the locations of the clusters in our sample strongly suggest that the region between the nuclei is the most obscured, and that the southern nucleus is more highly obscured than the northern nucleus.  This evidence suggests that deeper observations would result in additional cluster detections, particularly between the two nuclei.

\subsection{Summary}\label{Summary}

The circumnuclear locations of the clusters detected in these new, high-resolution observations strongly suggest that their formation was triggered by the merger event that created the ULIRG NGC 6240.  This motivates our decision to model the clusters as solar metallicity, instantaneous burst stellar populations.  If we further assume that the stellar populations were created with a Salpeter IMF, and that the stars in the clusters follow Padova evolutionary tracks, then we find that 27 of the 32 newly discovered clusters are consistent with being $\sim15$ Myr old, and 4 of the 32 clusters are probably between 3 and 15 Myr old.  Cluster ages much younger or much older than 15 Myr can be ruled out by the large star formation rates and masses inferred, respectively.  Assuming the clusters are 15 Myr old, then the majority of the clusters have masses ranging from $7 \times 10^{5} \rm M_{\odot}$ to $4 \times 10^{7} \rm M_{\odot}$, and the total contribution to the SFR from these 27 clusters is $\sim 10 \rm M_{\odot} \ yr^{-1}$, or roughly 10\% of the total SFR in the nuclear region.  

The extinctions calculated toward these clusters range from $\rm A_{V}=0.5$ -- 14 mag, with the least extinguished clusters being located farthest from the double nuclei.  While the range of extinctions quoted in the literature for NGC 6240 varies greatly, the values measured in this paper agree well with some of the larger published values.  The locations of the clusters detected here, along with the extinction map published by \citet{tec00}, suggest that there may be as-yet-unobserved clusters located between the nuclei that are still obscured.  The complicated morphology of this ULIRG makes it very difficult to determine our completeness limits empirically.

\acknowledgements
We would like to thank Peter Bodenheimer, Seran Gibbard, Raja Guhathakurta, David Koo, Jason Melbourne, Lynne Raschke, Constance Rockosi, and Jay Strader, for many helpful discussions.  The authors would also like to thank Gabriela Canalizo for reducing the NIRC2 K' data, and we would like to acknowledge the helpful support staff at Keck observatory.  Data presented herein were obtained at the W. M. Keck Observatory, which is operated as a scientific partnership among the California Institute of Technology, the University of California, and the National Aeronautics and Space Administration. The Observatory and the Keck II adaptive optics system were both made possible by the generous financial support of the W.M. Keck Foundation. The authors wish to extend special thanks to those of Hawaiian ancestry on whose sacred mountain we are privileged to be guests.  This work was supported in part by the National Science Foundation Science and Technology Center for Adaptive Optics, managed by the University of California at Santa Cruz under cooperative agreement No. AST-9876783.  This work was based, in part, on observations made with the NASA/ESA Hubble Space Telescope, obtained from the Data Archive at the Space Telescope Science Institute, which is operated by the Association of Universities for Research in Astronomy, Inc., under NASA contract NAS 5-26555.  GS acknowledges support provided by NASA through a grant from the Space  Telescope Science Institute, under NASA contract NAS 5-26555.  This work was also supported in part by NASA grant NAG 5-3042.

\clearpage
\begin{deluxetable}{lccc}
\tablecaption{\sc{Keck AO Observations of NGC 6240 using NIRC-2 Narrow Camera} \label{tableObsKeck}}
\tablewidth{0pt}
\tablehead{ \colhead{Date} & \colhead{Filter} & \colhead{Integration Time (s)} & \colhead{Guide Star's} \\ \colhead{} & \colhead{} & \colhead{} & \colhead{On-Axis Strehl}}
\startdata
24 Apr 2005 & H & 800 & 0.21 \\
17 Aug 2003 & K' & 630 & 0.32 \\
\enddata
\end{deluxetable}

\begin{deluxetable}{lcclc}
\tablecaption{\sc{Archival HST Observations of NGC 6240} \label{tableObsHST}}
\tablewidth{0pt}
\tablehead{ \colhead{Observing Program} & \colhead{Instrument} & \colhead{Camera} & \colhead{Filter} & \colhead{Exposure Time (s)} }
\startdata
GO 6430 (PI: van der Marel) & WFPC2 & PC & F814W (B) & $3\times 400$ \\
                                                    & WFPC2 & PC & F450W (I) & $3\times 700$ \\
GTO 7219 (PI: Scoville)         & NICMOS & NIC2 & F110W & $4\times 40$ \\
                                                   & NICMOS & NIC2 & F160W & $4\times 48$ \\
                                                   & NICMOS & NIC2 & F222M & $4\times 56$
\enddata
\end{deluxetable}

\clearpage
\clearpage
\begin{deluxetable}{cllcccc}
\tablecaption{\sc{Identification of Infrared Clusters in NGC 6240}\label{clusterTable}}
\rotate
\tablewidth{0pt}
\tablehead{\colhead{Cluster} & \colhead{$\Delta$ X} & \colhead{$\Delta$ Y} & \colhead{M$_{\rm V}$ (mag)} & \colhead{A$_{\rm V}$ (mag)} & \colhead{M$/(10^{6} \rm M_{\odot})$} & \colhead{SFR (M$_{\odot}$ yr$^{-1}$)} \\ \colhead{ID \#} & \colhead{(\arcsec E)} & \colhead{(\arcsec N)} & \colhead{2.9, 14.5, 180 Myr} & \colhead{2.9, 14.5, 180 Myr} & \colhead{2.9, 14.5, 180 Myr} & \colhead{2.9, 14.5, 180 Myr}}
\startdata
1 & 1.43 & 9.95 & -14.9, \textbf{-13.0}, -13.6 & 2.5, \textbf{0.5}, 1.0 & 2.8, \textbf{0.9}\tablenotemark{a}, 7.9 & 0.97, \textbf{0.06}, 0.04 \\ 
2 & -0.96 & 4.89 & -18.0, \textbf{-15.8}, -16.4 & 4.5, \textbf{2.0}, 2.5 & 49.0, \textbf{11.4}\tablenotemark{a}, 105.7 & 17.00, \textbf{0.79}, 0.59 \\ 
3 & -0.06 & 2.23 & -17.6, \textbf{-14.1}, -14.8 & 12.0, \textbf{6.5}, 7.0 & 34.0, \textbf{2.4}\tablenotemark{b}, 23.9 & 11.81, \textbf{0.17}, 0.13 \\ 
4 & -7.31 & 1.87 & -16.7, \textbf{-14.5}, -15.1 & 6.0, \textbf{3.5}, 4.0 & 14.8, \textbf{3.4}\tablenotemark{a}, 31.9 & 5.13, \textbf{0.24}, 0.18 \\ 
5 & 0.01 & 1.34 & -19.1, \textbf{-15.8}, -16.1 & 18.0, \textbf{13.5}, 11.0 & 137.5, \textbf{11.1}\tablenotemark{b}, 77.3 & 47.73, \textbf{0.77}, 0.43 \\ 
6 & 0.41 & 1.12 & \textbf{-17.7}, -14.4, -15.1 & \textbf{9.0}, 5.0, 6.0 & \textbf{35.8}, 3.0, 31.4 & \textbf{12.42}, 0.21, 0.17 \\ 
7 & -0.80 & 0.63 & -17.2, \textbf{-13.8}, -14.3 & 10.5, \textbf{6.0}, 5.0 & 22.9, \textbf{1.8}\tablenotemark{b}, 15.2 & 7.95, \textbf{0.13}, 0.08 \\ 
8 & -0.15 & 0.44 & -15.7, \textbf{-12.4}, -13.1 & 7.0, \textbf{3.5}, 4.0 & 5.6, \textbf{0.5}\tablenotemark{b}, 4.9 & 1.93, \textbf{0.03}, 0.03 \\ 
9 & -0.46 & 0.42 & -18.0, \textbf{-14.6}, -15.4 & 10.5, \textbf{6.0}, 7.5 & 48.8, \textbf{3.9}\tablenotemark{b}, 42.8 & 16.93, \textbf{0.27}, 0.24 \\ 
10 & -0.67 & 0.41 & -16.7, \textbf{-13.3}, -14.0 & 9.5, \textbf{5.0}, 5.5 & 13.8, \textbf{1.1}\tablenotemark{b}, 10.8 & 4.80, \textbf{0.08}, 0.06 \\ 
11 & -0.82 & 0.32 & -18.3, \textbf{-15.0}, -15.4 & 15.0, \textbf{11.0}, 9.0 & 63.5, \textbf{5.4}\tablenotemark{b}, 39.9 & 22.05, \textbf{0.37}, 0.22 \\ 
12 & 0.26 & 0.17 & -20.4, \textbf{-17.1}, -17.8 & 10.0, \textbf{6.5}, 6.5 & 433.5, \textbf{39.0}\tablenotemark{b}, 359.6 & 150.53, \textbf{2.70}, 1.99 \\ 
13 & -0.25 & 0.15 & -17.6, \textbf{-14.3}, -14.9 & 10.0, \textbf{5.5}, 6.0 & 34.6, \textbf{2.8}\tablenotemark{b}, 27.2 & 12.02, \textbf{0.19}, 0.15 \\ 
14 & 0.28 & 0.11 & -20.1, \textbf{-16.7}, -17.4 & 8.0, \textbf{3.5}, 4.0 & 325.2, \textbf{26.2}\tablenotemark{b}, 255.2 & 112.93, \textbf{1.81}, 1.41 \\ 
15 & 0.00 & 0.00 & -21.0, \textbf{-17.5}, -18.2 & 14.5, \textbf{9.0}, 9.5 & 782.1, \textbf{56.3}\tablenotemark{b}, 549.1 & 271.57, \textbf{3.90}, 3.04 \\ 
16 & -0.41 & -0.06 & -19.1, \textbf{-15.5}, -16.2 & 14.0, \textbf{8.0}, 8.5 & 129.7, \textbf{8.8}\tablenotemark{b}, 86.2 & 45.05, \textbf{0.61}, 0.48 \\ 
17 & -0.27 & -0.13 & -18.6, \textbf{-15.1}, -15.9 & 12.0, \textbf{6.5}, 7.5 & 86.2, \textbf{6.2}\tablenotemark{b}, 64.0 & 29.93, \textbf{0.43}, 0.35 \\ 
18 & -0.53 & -0.37 & -18.9, \textbf{-15.2}, -16.0 & 15.0, \textbf{8.5}, 10.0 & 105.9, \textbf{6.8}\tablenotemark{b}, 74.3 & 36.77, \textbf{0.47}, 0.41 \\ 
19 & 0.06 & -0.94 & -16.8, \textbf{-13.4}, -14.1 & 11.0, \textbf{6.0}, 6.5 & 16.4, \textbf{1.3}\tablenotemark{b}, 12.2 & 5.71, \textbf{0.09}, 0.07 \\ 
20 & 0.19 & -0.97 & -17.3, \textbf{-13.9}, -14.6 & 12.5, \textbf{7.5}, 8.0 & 26.1, \textbf{2.0}\tablenotemark{b}, 19.4 & 9.06, \textbf{0.14}, 0.11 \\ 
21 & -0.45 & -1.15 & -18.4, \textbf{-15.1}, -15.8 & 11.0, \textbf{6.5}, 7.5 & 71.6, \textbf{5.8}\tablenotemark{b}, 59.4 & 24.85, \textbf{0.40}, 0.33 \\ 
22 & -0.62 & -1.19 & -16.2, \textbf{-13.0}, -13.7 & 8.5, \textbf{5.0}, 6.0 & 9.3, \textbf{0.8}\tablenotemark{b}, 8.7 & 3.24, \textbf{0.06}, 0.05 \\ 
23 & -1.07 & -1.29 & -17.6, \textbf{-14.3}, -15.0 & 11.0, \textbf{7.0}, 7.0 & 34.7, \textbf{2.9}\tablenotemark{b}, 27.2 & 12.04, \textbf{0.20}, 0.15 \\ 
24 & 1.94 & -1.60 & -14.4, \textbf{-12.5}, -12.8 & 4.0, \textbf{2.0}, 2.0 & 1.8, \textbf{0.5}\tablenotemark{b}, 3.8 & 0.61, \textbf{0.04}, 0.02 \\ 
25 & 0.13 & -2.21 & -16.7, \textbf{-13.5}, -14.2 & 8.0, \textbf{5.0}, 5.5 & 14.5, \textbf{1.4}\tablenotemark{b}, 13.4 & 5.03, \textbf{0.10}, 0.07 \\ 
26 & -0.22 & -2.46 & -15.9, \textbf{-12.7}, -13.2 & 8.5, \textbf{5.5}, 4.5 & 7.2, \textbf{0.7}\tablenotemark{b}, 5.7 & 2.51, \textbf{0.05}, 0.03 \\ 
27 & 0.89 & -2.62 & -18.0, \textbf{-14.6}, -15.3 & 19.0, \textbf{14.0}, 15.0 & 50.0, \textbf{3.8}\tablenotemark{b}, 39.2 & 17.35, \textbf{0.26}, 0.22 \\ 
28 & -0.19 & -3.10 & \textbf{-17.9}, -14.6, -15.3 & \textbf{8.0}, 4.5, 5.5 & \textbf{42.3}, 3.8, 39.2 & \textbf{14.68}, 0.26, 0.22 \\ 
29 & 6.58 & -4.13 & \textbf{-17.0}, -14.5, -15.1 & \textbf{6.5}, 3.5, 4.0 & \textbf{19.3}, 3.4, 31.9 & \textbf{6.71}, 0.24, 0.18 \\ 
30 & 0.95 & -4.25 & -16.0, \textbf{-13.2}, -13.8 & 7.5, \textbf{4.0}, 4.5 & 7.3, \textbf{1.0}\tablenotemark{b}, 9.2 & 2.53, \textbf{0.07}, 0.05 \\ 
31 & -1.02 & -5.30 & -14.4, \textbf{-12.8}, -13.4 & 3.5, \textbf{2.0}, 2.5 & 1.8, \textbf{0.7}\tablenotemark{a}, 6.6 & 0.62, \textbf{0.05}, 0.04 \\ 
32 & -3.54 & -7.71 & \textbf{-19.0}, -16.2, -16.9 & \textbf{7.5}, 4.0, 4.5 & \textbf{124.7}, 17.0, 157.6 & \textbf{43.31}, 1.18, 0.87 \\ 
\enddata
\tablecomments{Column 1 specifies each cluster using the identification numbers as shown in Figure~\ref{ClusterIDs}.  Smaller numbers correspond to more northern clusters.  Columns 2 and 3 give the clusters' positions relative to cluster \#15, which is coincident with the site of the northern black hole.  Column 4 shows the unextincted V band absolute magnitude each cluster would have if it were 2.9, 14.5, and 180 Myr old, and if it were extincted by an amount specified in column 5. (See text for detailed description of the assumptions made for the stellar populations.)  Column 5 specifies the minimum amount of extinction needed for the measured SEDs to match the model SEDs for the 3 cluster ages. Column 6 gives the mass of each cluster for the 3 cluster ages assuming the values of $\rm A_{V}$ listed in column 5.  Errors on the mass and extinction can be deduced from the error bars shown in Figures~\ref{seds_7_18}-\ref{seds_1_32}.  Column 7 shows the star formation rate implied by the masses and ages listed in column 6; we simply divide the mass by the age in each case.  Bold text indicates which of the three stellar population ages best agrees with our observations, as determined through visual inspection.}
\tablenotetext{a}{Using visual inspection of model SEDs, it is difficult to distinguish between the 14.5 and 180 Myr-old stellar populations, while we can rule out the 3 Myr population.  In these cases, to be consistent with our method of quoting the minimum amount of extinction needed for the measured SEDs to match the model SEDs, we choose as the most likely cluster age that which yields the smallest cluster masses.  The 14.5 Myr-old stellar population implies a less massive cluster.}
\tablenotetext{b}{Using visual inspection of model SEDs, there is insufficient data to distinguish between stellar populations with ages ranging from 2.9 to 180 Myr.  In these cases we choose as the most likely cluster age that which yields the smallest cluster masses.  The 14.5 Myr-old stellar population corresponds to the least massive cluster.}
\end{deluxetable}

\clearpage

\clearpage

\clearpage

\clearpage

\clearpage

\clearpage

\clearpage

\clearpage

\clearpage

\end{document}